\documentclass[amsmath,amssymb,aps,prx,notitlepage]{revtex4-1}
\usepackage{amssymb}
\usepackage{amsmath}
\usepackage[utf8]{inputenc}
\usepackage{graphicx}
\usepackage{dsfont}
\usepackage{comment}
\usepackage{xcolor}
\def\Tr{\mathrm{Tr}}
\newcommand{\red}[1]{\textcolor{black}{#1}}

\begin{document}
\title{Benchmarking photon number resolving detectors}
\author{Jan Provazn\'ik}
\author{Luk\'a\v{s} Lachman}
\author{Radim Filip}
\author{Petr Marek}
\affiliation{Department of Optics, Palack\'y University, 17. listopadu 1192/12, 771 46 Olomouc, Czech Republic}

%\address{Department of Optics, Palacky University, 17. listopadu 1192/12, 77146 Olomouc, Czech Republic}
%\email{\authormark{*}marek@optics.upol.cz}

\begin{abstract}
Photon number resolving detectors are the ultimate measurement of quantum optics, which is the reason why developing the technology is getting significant attention in recent years. With this arises the question of how to evaluate the performance of the detectors. We suggest that performance of a photon number detector can be evaluated by comparing it to a multiplex of on-off detectors in a practical scenario: conditional preparation of a photon number state. Here, both the quality of the prepared state and the probability of the preparation are limited by the number of on-off detectors in the multiplex, which allows us to set benchmarks that can be achieved or surpassed by the photon number resolving detectors.
\end{abstract}
\maketitle
	
\section{Introduction}
Photon number resolving detectors are \emph{the} most sought measurement of quantum optics. By counting the individual photons of the signal they are the ultimate refinement of intensity detectors, which are currently the dominant way of observing light. Even the trademarks of modern quantum optics, the homodyne detectors capable of directly measuring the electric intensity of the field, rely on continuous detection of intensity of light.
Photon number resolving detectors (PNRD) are also crucial from the perspective of contemporary quantum optics and its focus on the transition from Gaussian to non-Gaussian resources. The non-Gaussian resources, be it states, operations, or measurements, are essential or at least advantageous for many applications, such as quantum computation \cite{KLM,PhysRevA.68.042319}, quantum metrology \cite{npjQI.2.16023, PhysRevA.100.013814, von_Helversen_2019}, or quantum communication \cite{NatPhot.9.48, npjQI.5.65}. PNRDs will unambiguously discriminate between states with different photon numbers and thus can be used for non-Gaussian quantum state preparation \cite{tiedau2019scalability} as well as for deterministic implementation of important non-Gaussian operations \cite{PhysRevA.64.012310, SandersCubic,PhysRevA.95.052352}.

The first successful attempts to detect individual photons came in the form of avalanche photodiodes capable of discerning presence of at least one photon in the field. These on-off detectors were instrumental for preparing isolated photons which in turn enabled the entire field of quantum optics with individual photons \cite{RevModPhys.84.777}. The avalanche photodiodes were also used to develop the first approximations of photon number resolving detectors based on multiplexing the optical field into separate spatial or temporal modes and counting separate detection events. Such detectors, which we will call Multiplexed Single Photon Detectors (MSPD), can be used for measuring photon number distributions \cite{PhysRevLett.97.043602, PhysRevLett.101.053601, PhysRevA.85.023816} and point values of Wigner functions \cite{Laiho_2009, PhysRevLett.105.253603, PhysRevLett.120.063607, PhysRevLett.116.133601}, and they were even used in preparation of higher order non-Gaussian states \cite{Yukawa:13} and implementation of non-Gaussian operations \cite{PhysRevA.81.022302,NoisePowered,PhysRevA.86.010305}. The quality of these detectors depends on their architecture as well as on the number of individual on-off detectors \cite{PhysRevA.99.043822}.

The most apparent feature of the MSPDs arises from the physical limit of $M$ individual detectors: the detector cannot detect photon numbers above this value. Furthermore, the probability of receiving an erroneous result increases with the number of photons in the detected signal. This would not be the case for an ideal PNRD that implements perfect projections on individual Fock states $|n\rangle$. Such the detector is an idealized construct, but it can be asymptotically achieved. The approach which, at the present time holds most promise for achieving it, employs superconducting transition edge sensors \cite{OE.16.3032, OE.Calkins.13,NatPhot.7.210,PhysRevLett.116.143601,PhysRevA.95.053806, PhysRevLett.118.163602}, which, instead of using a number of individual detectors, take advantage of physical processes that have a measurable inherent response to extremely low changes of intensity of the incident light.

However, at the current state of the art, the ideal photon number resolving capability remains a distant goal for both MSPDs and realistic PNRDs. This comes with an interesting question, though: how can we quantify how far from the goal are we? While a natural parameter for MSPDs is the number of detectors used in their composition, for PNRDs the situation is more complicated. In this paper we suggest that even PNRDs can be evaluated in this way. We show that for conditional preparation of photon number states, which is a practical task that tests both the nature of the detectors and their integrability with the standard tools of quantum optics \cite{NatPh.Andersen.15}, there is no qualitative difference between imperfect PNRDs and MSPDs with a specific number of detectors. Consequently, performance of PNRDs can be quantified by the number of detectors forming a MSPD that provides equivalent results.

\section{State preparation by detection}
\begin{figure}
    \centering
\includegraphics[width=0.6\textwidth]{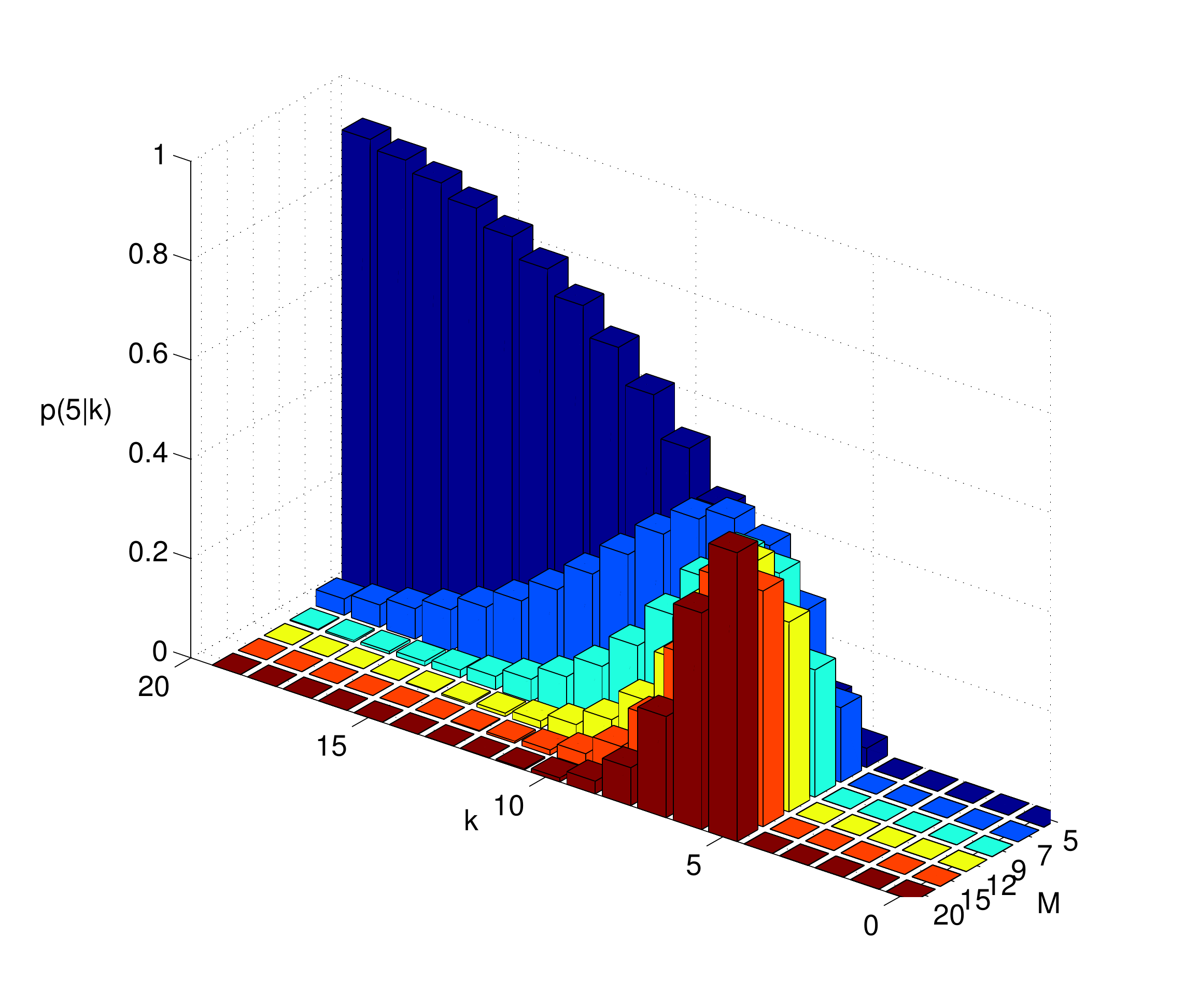}
    \caption{POVM elements for detecting exactly five photons by  MSPD  composed of $M$ individual APD detectors. For comparison, Ideal PNR detector has $p(k|5) = \delta_{k5}$. }
    \label{fig_POVMAPD}
\end{figure}

\begin{figure}
    \centering
\includegraphics[width=0.6\textwidth]{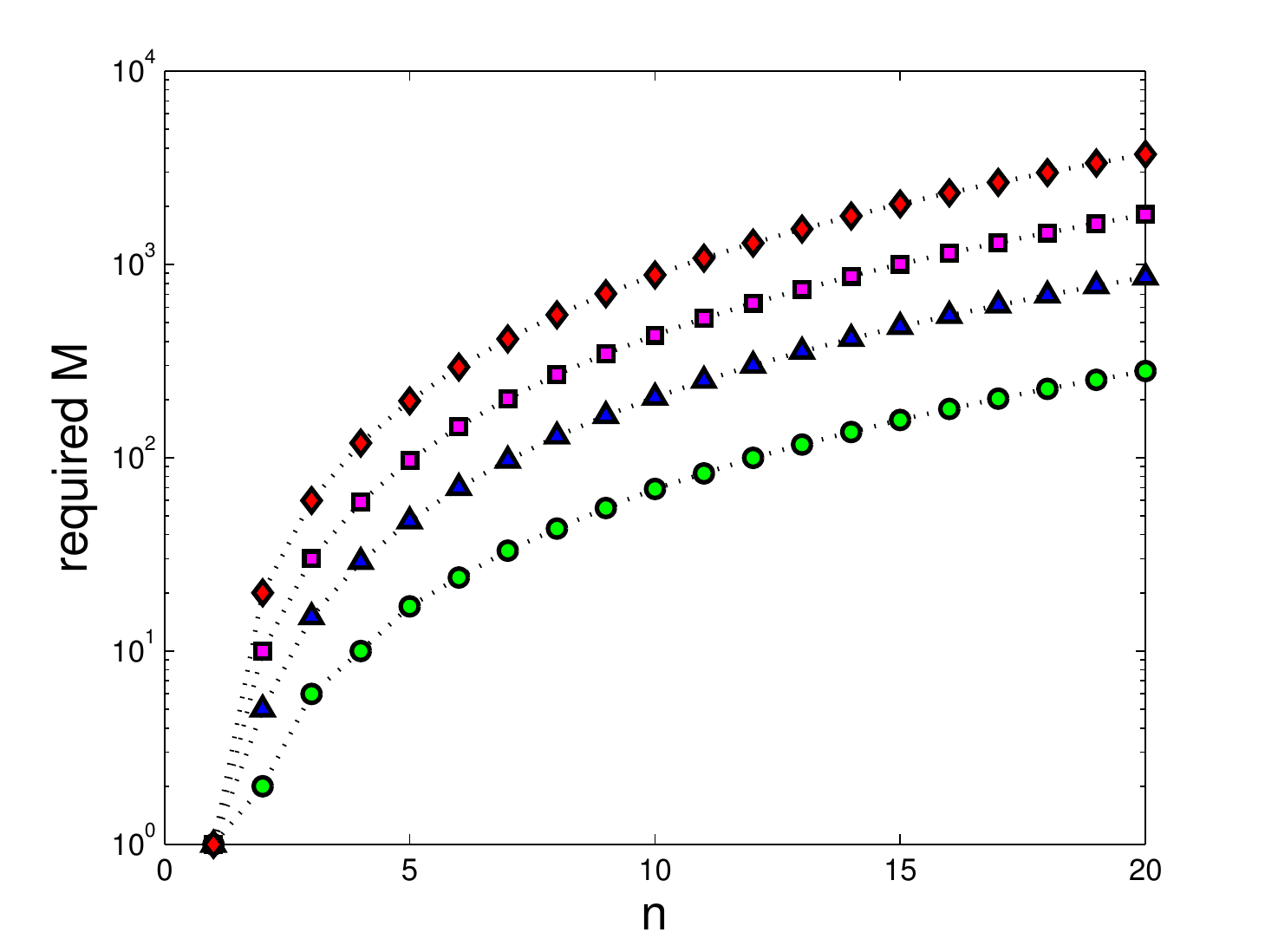}
    \caption{The number of detectors $M$ required for $p(n|n)$ larger than a certain threshold: green diamonds - 0.5, blue triangles - 0.8, magenta squares - 0.9, red diamonds - 0.95. }
    \label{fig_MrequiredforPNN}
\end{figure}

In general, quantum physics represents detectors by positive operator valued measures (POVM). For each possible measurement outcome, there is an non-negative hermitian operator POVM element which can be used to determine the probability of this outcome and its action on the rest of the physical system. For the ideal single mode PNRD, these POVM elements are simple. An outcome related to observing $n$ photons has POVM element which is just projector on the respective Fock state, $\Pi_{PNRD} = |n\rangle\langle n|$. For ideal MSPDs, which are constructed by multiplexing the measured mode into $M$ spatial or temporal modes, measuring each of them by a single on-off detector, and taking $n$ as the number of positive events, the POVM can be derived to be:
\begin{equation}\label{POVM_APD}
    \Pi_{MSPD} = \sum_{k=0}^{\infty} p(n|k) |k\rangle\langle k|,
\end{equation}
where the conditional probabilities are
\begin{equation}\label{POVM_APDprob}
 p(n|k) = \frac{1}{M^k} \left( \begin{array}{c}
  M \\
  n
       \end{array}\right)
    \sum_{l= 0}^{n-1} \left( \begin{array}{c}
                               n \\
                               l
                             \end{array}\right)
                             (-1)^l (n-l)^k
\end{equation}
when $k \le n \le M $, and zero otherwise \cite{PhysRevLett.76.2464}. An example of POVM elements for projecting on state $|5\rangle$ is shown in  Fig.~\ref{fig_POVMAPD}. We can see that MSPDs can can provide false positives which appear when two or more photons imping on a single detector that can not differentiate them. This effect diminishes with the number of detectors $M$, but the rate of this scales almost exponentially with the number of detectors, $M$, required. This can be seen on another example, shown in Fig.~\ref{fig_MrequiredforPNN}, where we plot the probabilities that photon number state  $|n\rangle$ produces detection event $n$, $p(n|n)$. In the limit of sufficiently large $M$, MSPDs become equivalent to PNRD (see the appendix for proof). This also means that for any imperfect PNRD, there will be $M$ such that ideal MSPD with this $M$ provides identical or better performance. This $M$ can be then taken as a figure of quality of the PNRD. And a practical scenario in which we can realize the comparison is the conditional preparation of photon number states.
\begin{figure}
    \centering
\includegraphics[width=0.7\textwidth]{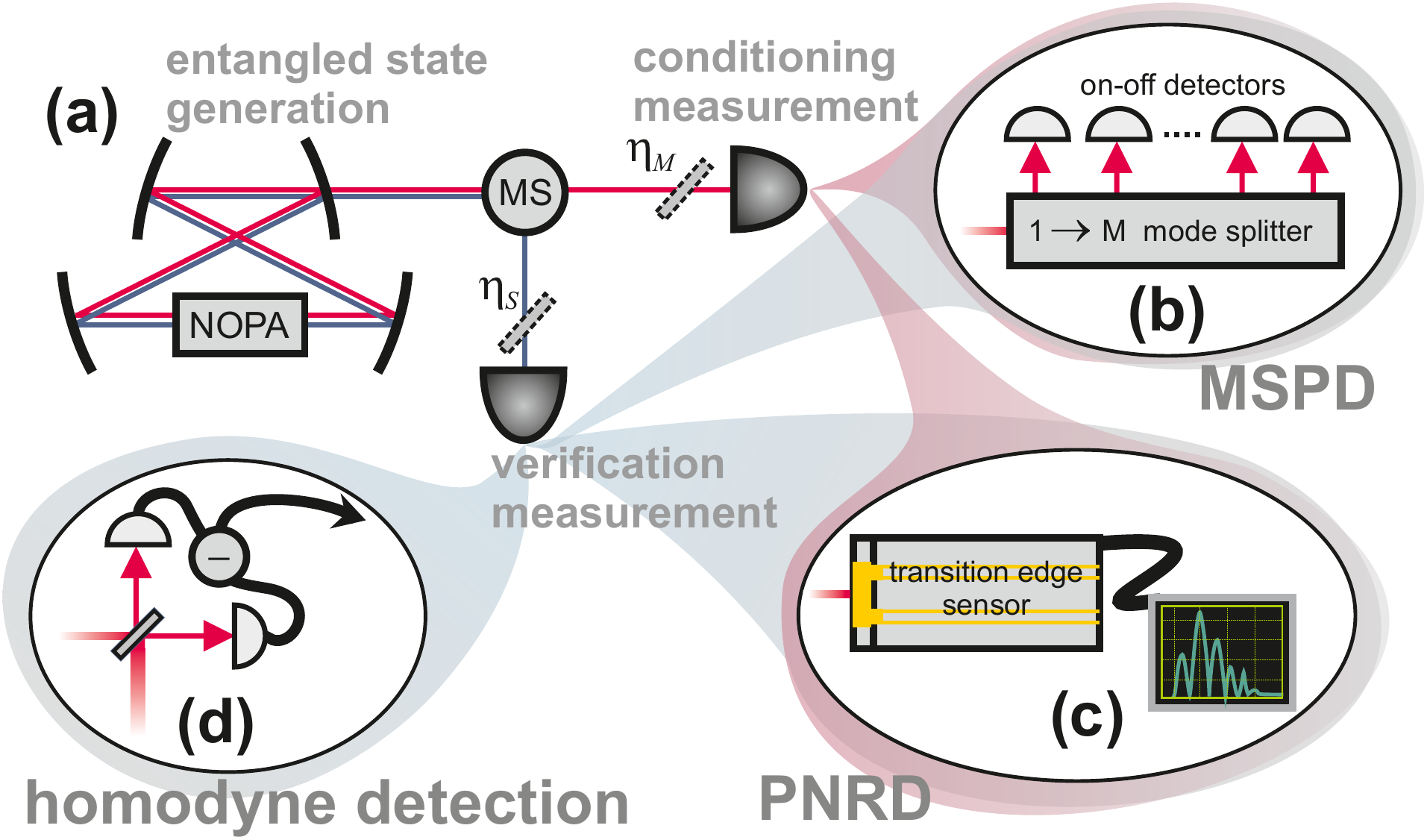}
    \caption{(a) Schematic depiction of experiment for preparation and verification of photon number states. Two modes of the entangled two-mode squeezed states are separated by mode separator (MS) and one of the modes is fed into the conditioning measurement realized by MSPD (b) or PNRD (c). The properties of the state thus prepared in the remaining mode can be verified by MSPD, PNRD, or homodyne detection (d). Virtual beam splitters with transmission coefficients $\eta_M$ and $\eta_S$ represent effective quantum efficiencies of the detectors.  }
    \label{fig_setup}
\end{figure}


Photon number states of light can be prepared by the setup sketched in Fig.~\ref{fig_setup}a by measuring one mode of a Gaussian two-mode squeezed state by a detector capable of discerning different Fock states and conditioning the remaining system when the desired outcome has been detected. Formally we can represent generated quantum state in mode $S$ as:
\begin{equation}\label{rhoS}
    \rho_{S} = \frac{1}{P_S}\Tr_M[ \rho_{S,M} \Pi_M],
\end{equation}
where $P_S = \Tr[ \rho_{S,M} \Pi_M]$ is the success rate,  $\rho_{S,M}$ is the state of the initial two-mode squeezed state, and $\Pi_M$ is the positive operator valued measure (POVM) element corresponding to the desired outcome of measuring modes $M$. In the ideal scenario, the two-mode squeezed state is pure and can be represented by state vector $|\psi_{TMS}\rangle_{S,M} = \sqrt{1-\lambda^2}\sum_{n = }^{\infty} \lambda^n |n,n\rangle_{S,M}$. In practice, various experimental inefficiencies will manifest as losses causing the resource sate to be mixed. The losses, separate for each of the subsystems, can be represented with help of the decoherence map
\begin{equation}\label{losses}
    \mathcal{D}_{j,\eta}(\rho) = \sum_{n = 0}^{\infty} \frac{(1-\eta)^n}{n!} \sqrt{\eta}^{a_j^{\dag}a_j}a_j^n\rho a_j^{\dag n}  \sqrt{\eta}^{a_j^{\dag}a_j},
\end{equation}
where $j = S,M$ denotes the modes and $a_j$ is the annihilation operator acting on them. A general mixed resource state can be then expressed as $\rho_{S,M} = \mathcal{D}_{S,\eta_S}[ \mathcal{D}_{M,\eta_M}(|\psi_{TMS}\rangle_{S,M}\langle \psi_{TMS}|)]$. The total detection and collection efficiency $\eta_M$ covers internal losses of preparing the two-mode squeezed state as well as coupling losses and quantum efficiency of the projecting detector. The system efficiency $\eta_S$ then mostly manifests as losses in the already produced quantum state and inefficiency of detectors used for its verification.

Without any losses, $\eta_M = \eta_S = 1$, PNRD always produces the perfect state $|n\rangle$ with probability $P_S = (1-\lambda^2)\lambda^{2n}$, which can be maximized with respect to the entanglement of the shared state parameterized by $\lambda$. On the other hand, MSPD can achieve the same Fock state only in the limit of large enough number of detectors $M \rightarrow \infty$ or in the limit of negligible entanglement $\lambda \rightarrow 0$. For a realistic number of detectors, MSPDs will always need to make concessions between high probability of success and high quality of the state.

\begin{figure}
    \centering
    \includegraphics[width=0.45\textwidth]{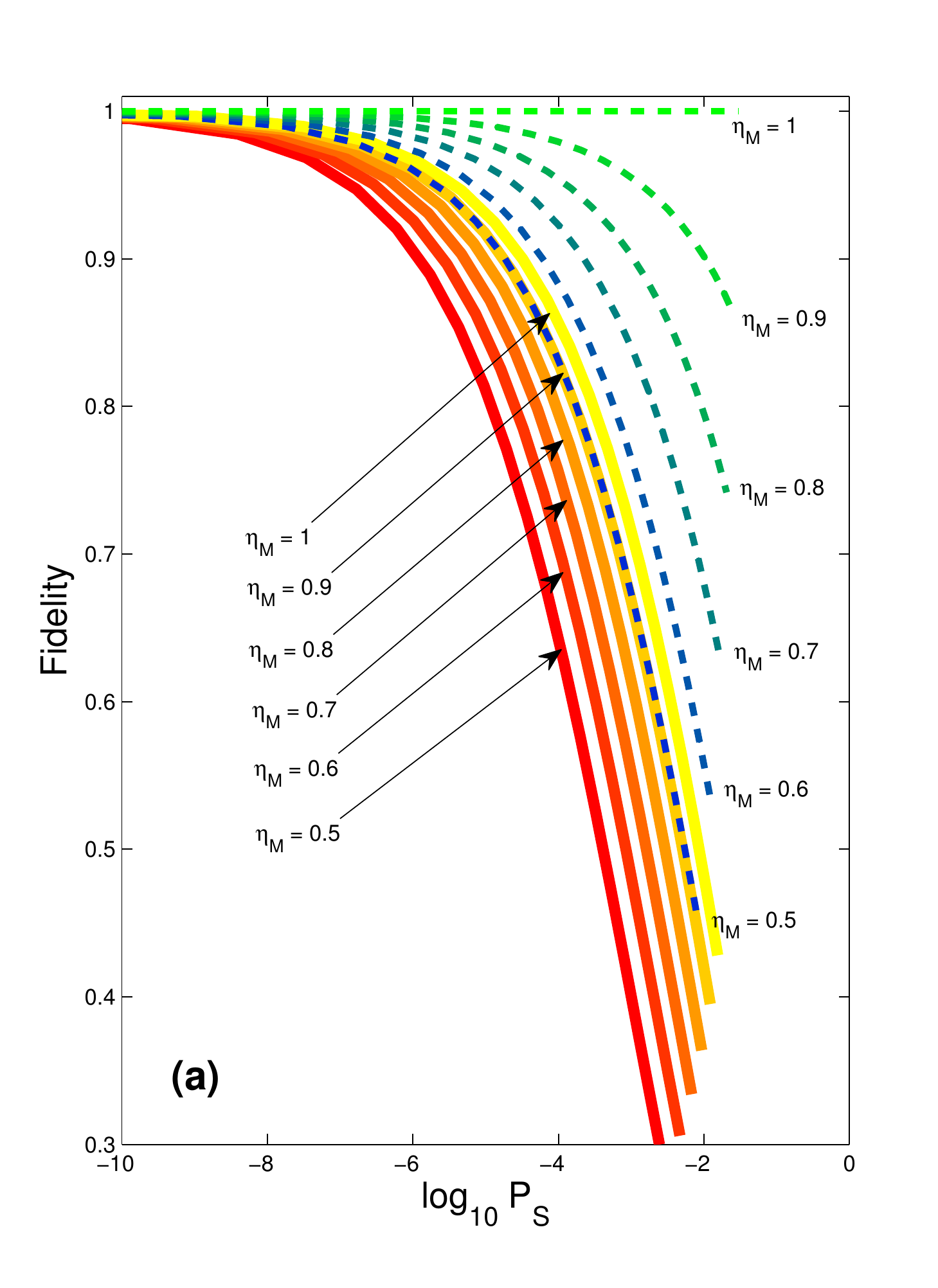}
    \includegraphics[width=0.45\textwidth]{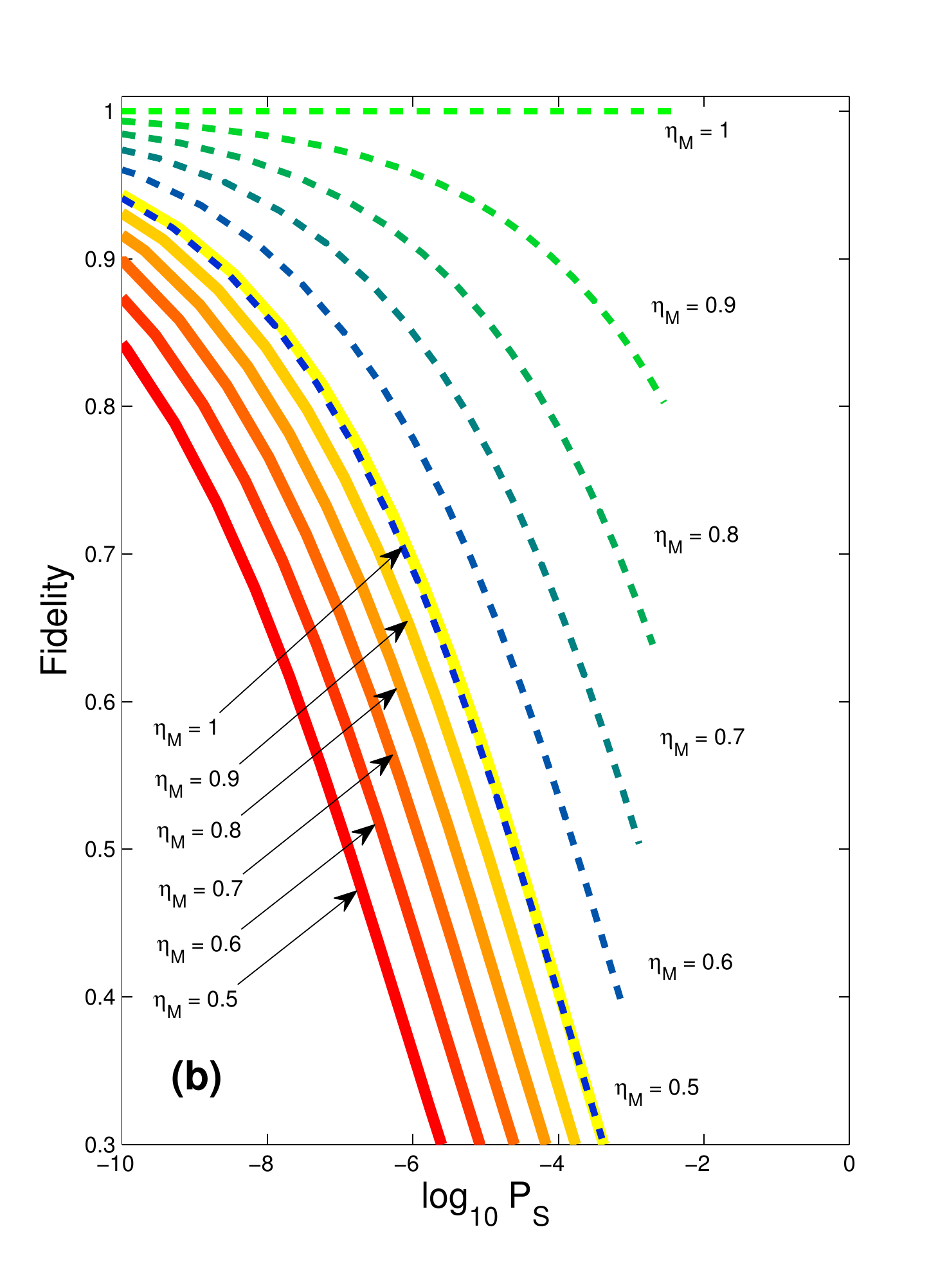}
    \caption{\red{Fidelity relative to success probability for preparation of Fock states $|3\rangle$ (a), and $|5\rangle$ (b). Dashed lines represent PNRD detectors with colors ranging from green ($\eta_M = 1$) to blue ($\eta_M = 0.5$). Solid lines represent MSPD detectors with $M=n$ with colors ranging from yellow ($\eta_M = 1$) to red ($\eta_M = 0.5$). The respective efficiencies for all lines are also marked in the figures. }}
    \label{fig_Fid_PS}
\end{figure}

The situation is not as clearly cut in the practical scenario in which the detectors are no longer perfect, $\eta_M \neq \eta_S = 1$, as in this case even PNRD cannot always produce state with perfect quality. In this case we can characterize the produced quantum state by two main parameters: the success rate $P_S = \Tr[ \rho_{S,M} \Pi_M]$ and the fidelity $F = \langle n| \rho_S| n \rangle$. For the case of preparing quantum states $|3\rangle$ and $|5\rangle$, the comparison of the detectors is shown in Fig.~\ref{fig_Fid_PS}. We can see that while both detectors show a trade-off between the fidelity and the probability of success, the PNRD, even with lower efficiency, still perform significantly better than even the ideal version of the sample MSPD. That is, MSPD can still generate states with fidelity as large as that of the states produced by PNRD, but their probability of success is significantly lower.

However, fidelity is not the only measure of quality of quantum states. Photon number states states are, by their very nature, highly nonclassical and non-Gaussian states. For odd photon number states fidelity higher than $F > 0.5$ guarantees negative Wigner function and thus both nonclassicality and nongaussianity, but using a witness with more general application, such as \cite{NPJ.Straka} is more prudent. \red{ To this end we apply hierarchy of witnesses of genuine quantum non-Gaussianity that has recently been derived and experimentally tested \cite{PhysRevLett.123.043601}. In the hierarchy, witness of $n$-th order discerns whether the investigated state is truly incompatible with a state that could be created as a mixture of photon number superposition states $\sum_{k=0}^{n-1}c_k|k\rangle$ transformed by arbitrary Gaussian operations.}

\begin{figure}
    \centering
    \includegraphics[width=0.45\textwidth]{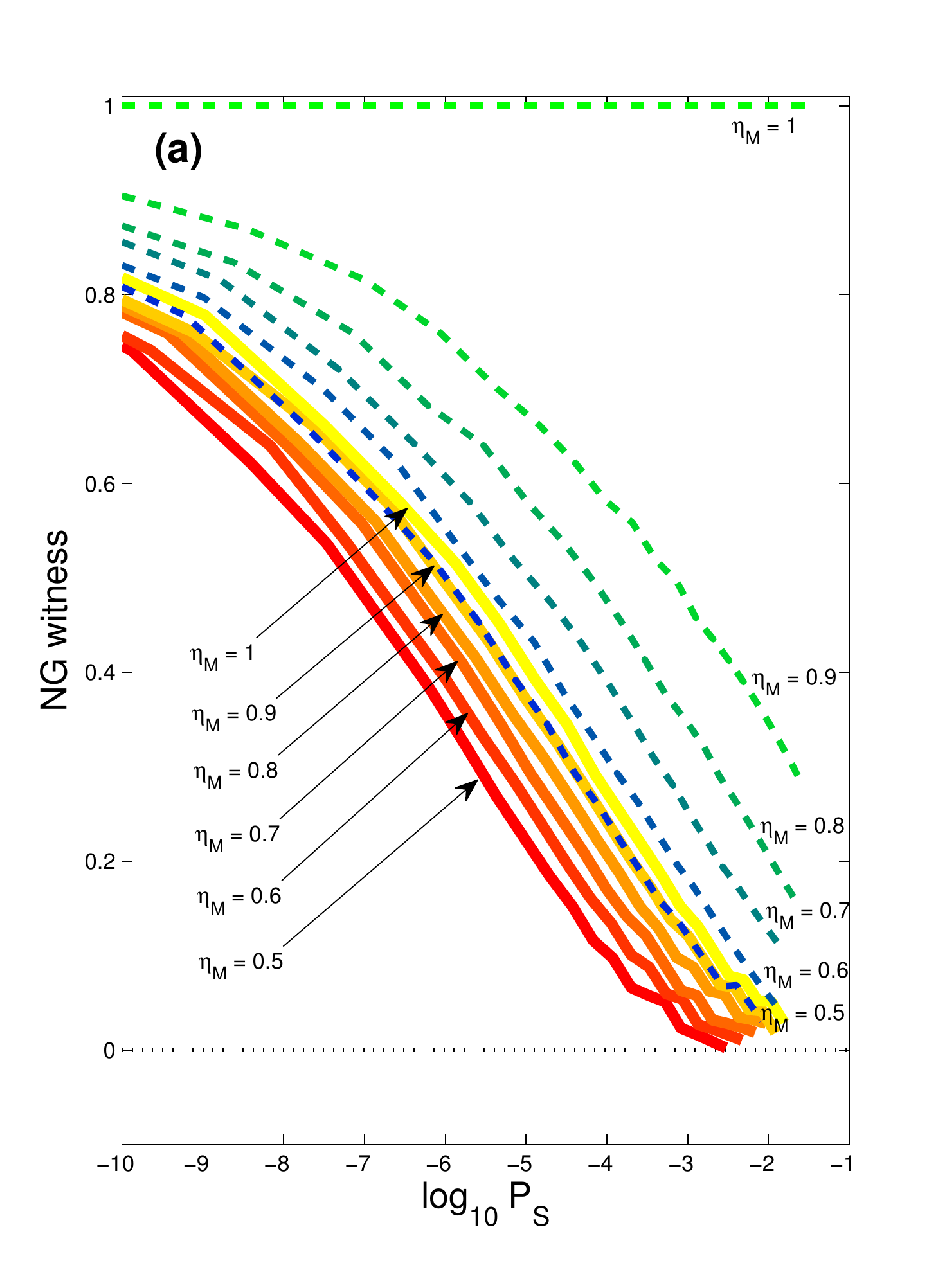}
    \includegraphics[width=0.45\textwidth]{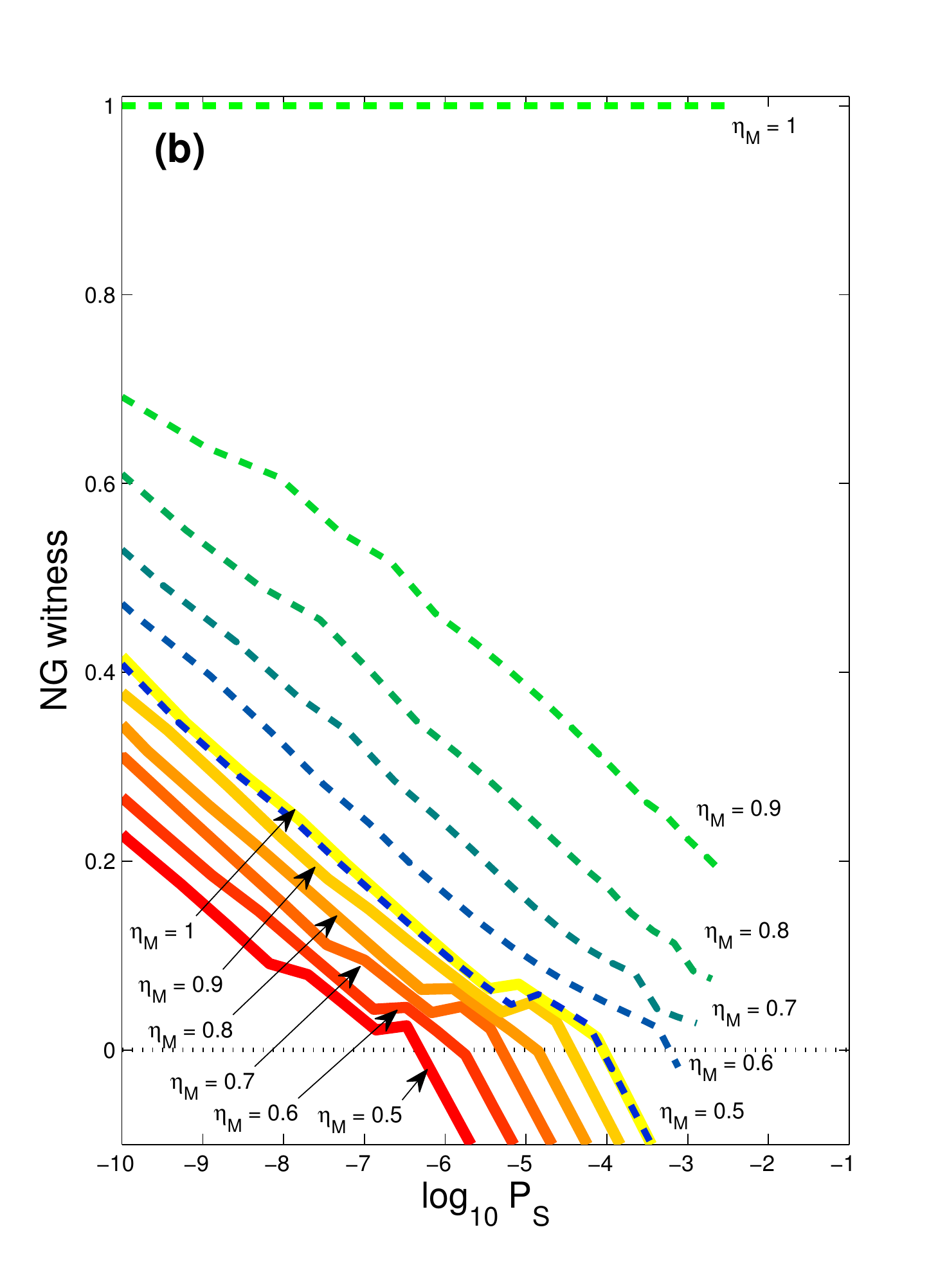}
    \caption{\red{Genuine quantum nongaussianity witnesses $W_{NG}(3)$ for Fock states $|3\rangle$ (a) and $W_{NG}(5)$ for  $|5\rangle$ (b). Dashed lines represent PNRD detectors with colors ranging from green ($\eta_M = 1$) to blue ($\eta_M = 0.5$). Solid lines represent MSPD detectors with $M=n$ with colors ranging from yellow ($\eta_M = 1$) to red ($\eta_M = 0.5$). The respective efficiencies for all lines are also marked in the figures. Genuine quantum non-Gaussian states have values of the witness above zero, which is marked by the black dotted line.}  }
    \label{fig_NG_PS}
\end{figure}

\red{One of the telling features of Gaussian states is that, with the exception of the vacuum state, their photon number representation spans an infinite dimensional Hilbert space. As a consequence, for Gaussian states the probability of detecting any particular photon number state $|n\rangle$, which can be represented by fidelity  $ F_n = \langle n|\rho_S|n\rangle$, is always bounded from above \cite{PhysRevLett.106.200401}. This remains true even if, instead of Gaussian states, we consider an arbitrary finite superpositions of photon number states $|k\rangle$ with $k<n$, which are affected by arbitrary Gaussian operations. This upper bound can be further made tighter by deriving it under specific constraints. One such constraint is limiting the cumulative probability of detecting photon number states with numbers larger than $n$, $p_{n+} = \Tr[ \rho_S \sum_{k=n+1}^{\infty}|k\rangle\langle k|]$. Value $B_n(p_{n+})$ can be then defined as the maximal fidelity $F_n$ that can be, for the specific value of $p_{n+}$, achieved by state $\sum_{k=0}^{n-1}c_k|k\rangle$ after it undergoes an arbitrary Gaussian operation \cite{PhysRevLett.123.043601}. The values of the bound can be numerically obtained and compared to the fidelities of the prepared states. As a numerical quantifier for the witness of $n$-th order
we can take the difference $W_{NG}(n) = F_n - B_n(p_{n+})$, where $F_n$ is the fidelity of the observed state, $p_{n+}$ is calculated from the photon number distribution of the observed state, and $B_n(p_{n+})$ is obtained by numerical optimization for the particular value \cite{PhysRevLett.123.043601}. If this value is positive it signifies that the state could not have been created from states with purely lower photon number components, even with help of arbitrary Gaussian operations. Such state is therefore genuinely $n$-photon quantum non-Gaussian state according to the definition in \cite{PhysRevLett.123.043601}. The values of the witnesses with appropriate orders are shown in Fig.~\ref{fig_NG_PS}. }
We can see that both kinds of detectors are capable of producing genuine quantum non-Gaussian states, both show similar trade-offs, and, again, PNRD visibly outperforms the MSPD. \red{This is most prominent for the displayed scenario with $M=n$. Increasing $M$ would move the MSPD curves towards PNRD curves and eventually, for $M\rightarrow \infty$, they would merge.}  

\begin{figure}
    \centering
\includegraphics[width=0.45\textwidth]{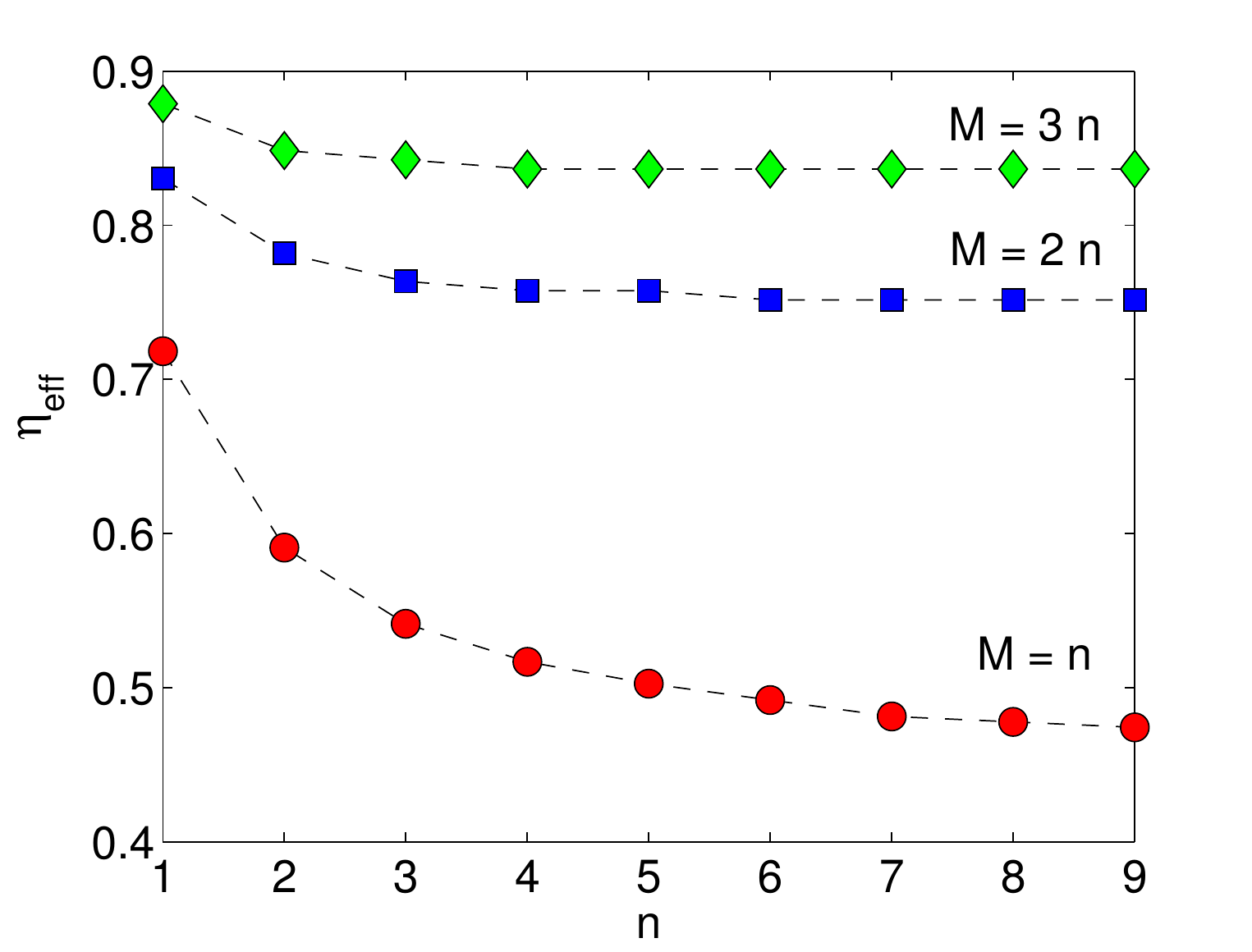}
    \caption{Effective efficiency with which imperfect PNRD operates at the level of ideal MSPDs for preparation of fock states $|n\rangle$. The total numbers of APDs forming the detectors are $M = n$ (red circles), $M = 2n$ (blue squares), and $M = 3n$ (green diamonds).  }
    \label{fig_eta_eff}
\end{figure}

Comparison of Fig.~\ref{fig_Fid_PS} and Fig.~\ref{fig_NG_PS} reveals strong similarity in the trends of the trade-offs with respect to detection efficiencies $\eta_M$. In both situations, photon number state $|5\rangle$ is more difficult to prepare than Fock state $|3\rangle$, meaning that the same quality is achieved at the cost of lower success rate. The minimal efficiency of PNRD required to overcome perfect MSPD also increases. Interestingly, in between the two figures of merit, the actual values of efficiency in which imperfect PNRD performs as perfect MSPD are equal. These efficiencies are are $\eta_M = 0.54$ for $M = 3$ and three photon state, and $M = 5$ and $\eta_M = 0.50$ for five photon state.
For each MSPD we can therefore define effective collection efficiency $\eta_{M,\mathrm{eff}}$ as the efficiency of PNRD leading to effectively identical trade-off between the fidelity of a prepared Fock state and the success probability of the procedure. These effective efficiencies for some composite detectors are shown in Fig.~\ref{fig_eta_eff}. These values start high and diminish with increasing $n$ as the performance of MSPDs worsens and PNRD with lower efficiency become competitive enough. The effective efficiencies depend on the ratio between the number of the detectors and the required $n$ of the Fock state. Numerical analysis reveals that the effective efficiencies approach $\eta_{M,\mathrm{eff}} \rightarrow 0.47$ for when the number of detectors is equal to the number of detected photons, then $\eta_{M,\mathrm{eff}} \rightarrow  0.75$ when the number of detectors is two-times higher and $\eta_{M,\mathrm{eff}}\rightarrow  0.84$ when it is three times higher. We can thus see that the number of detectors forming a MSPD can play a similar role as efficiency of PNRD.

\section{Benchmarking tools}
We have established that for preparation of a specific photon number states imperfect PNRDs can behave similarly to MSPDs with some number of ideal detectors $M$. We can use this to experimentally characterize the quality of an unknown PNRD. The experiment, sketched in Fig.~\ref{fig_setup}, attempts to conditionally prepare the photon number state by suitable detection. We can always calculate the performance of MSPDs composed of $M$ on-off detectors and compare it to the experimental results.
\begin{figure}
    \centering
\includegraphics[width=0.45\textwidth]{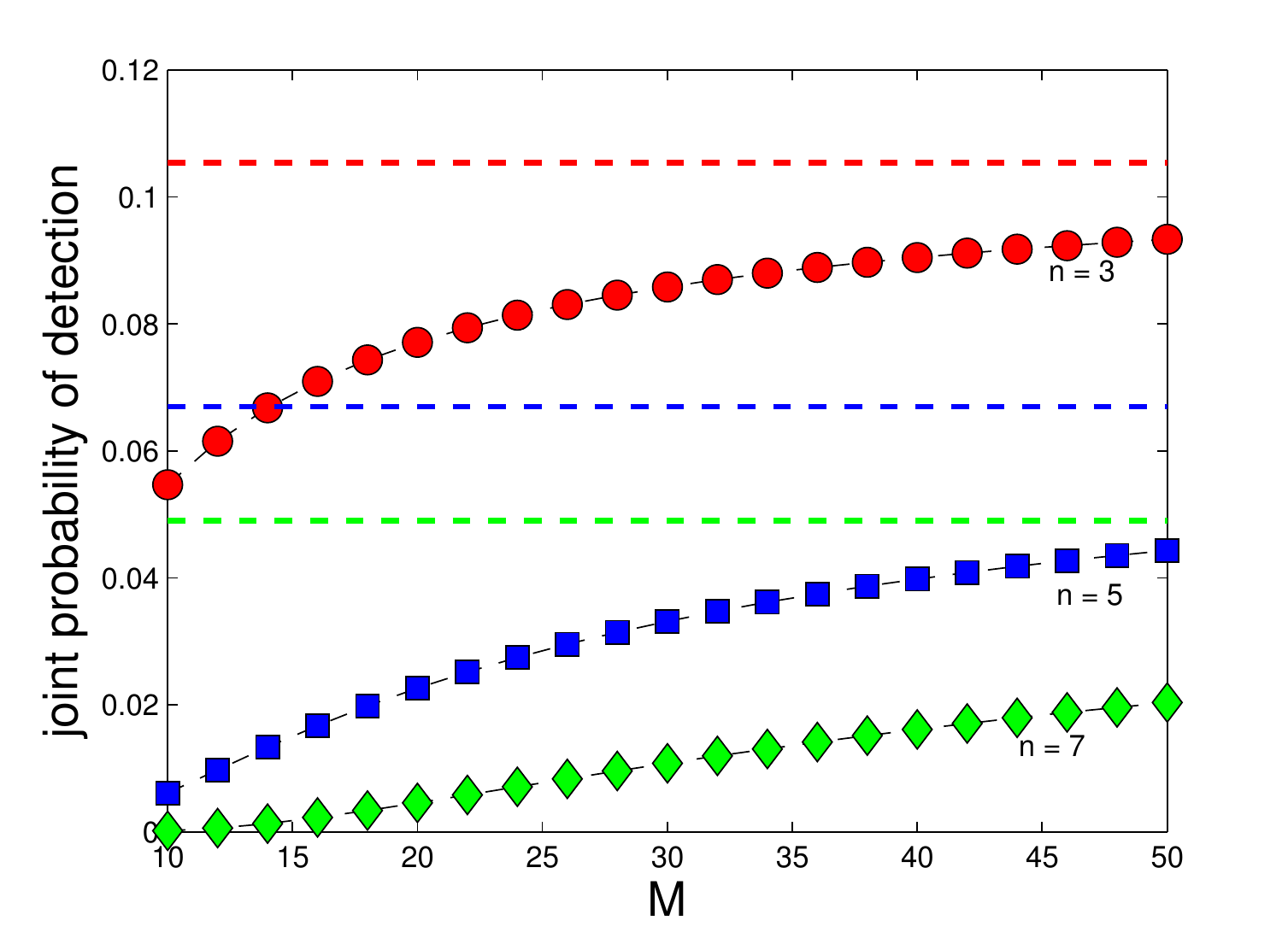}
    \caption{Joint probability benchmark for simultaneous detection of fock state $|3\rangle$ (red circles), $|5\rangle$ (blue squares), and $|7\rangle$ (green diamonds) by two ideal MSPDs with variable $M$. The dashed lines with the respective colors represent maximal joint probability achievable by perfect PNRD. }
    \label{fig_jointP}
\end{figure}
In the first step we can consider using the same detector for both preparation and subsequent verification. In this case the suitable figure of merit is the joint probability of detecting a selected photon number state $|n\rangle$, which can be obtained as $P_{nn} = \max_{\lambda}\Tr[\rho_S \Pi_M\otimes\Pi_M]$, where the POVM elements $\Pi_M$ correspond to the desired detection result. Fig.~\ref{fig_jointP} shows the maximal joint probabilities that can be, for MSPD with given number of detectors $M$, obtained when preparing photon number states $|3\rangle$, $|5\rangle$, and $|7\rangle$. Experimentally obtaining higher joint probability then guarantees that detector performs better than the respective PNRD.

\begin{figure}
    \centering
\includegraphics[width=0.45\textwidth]{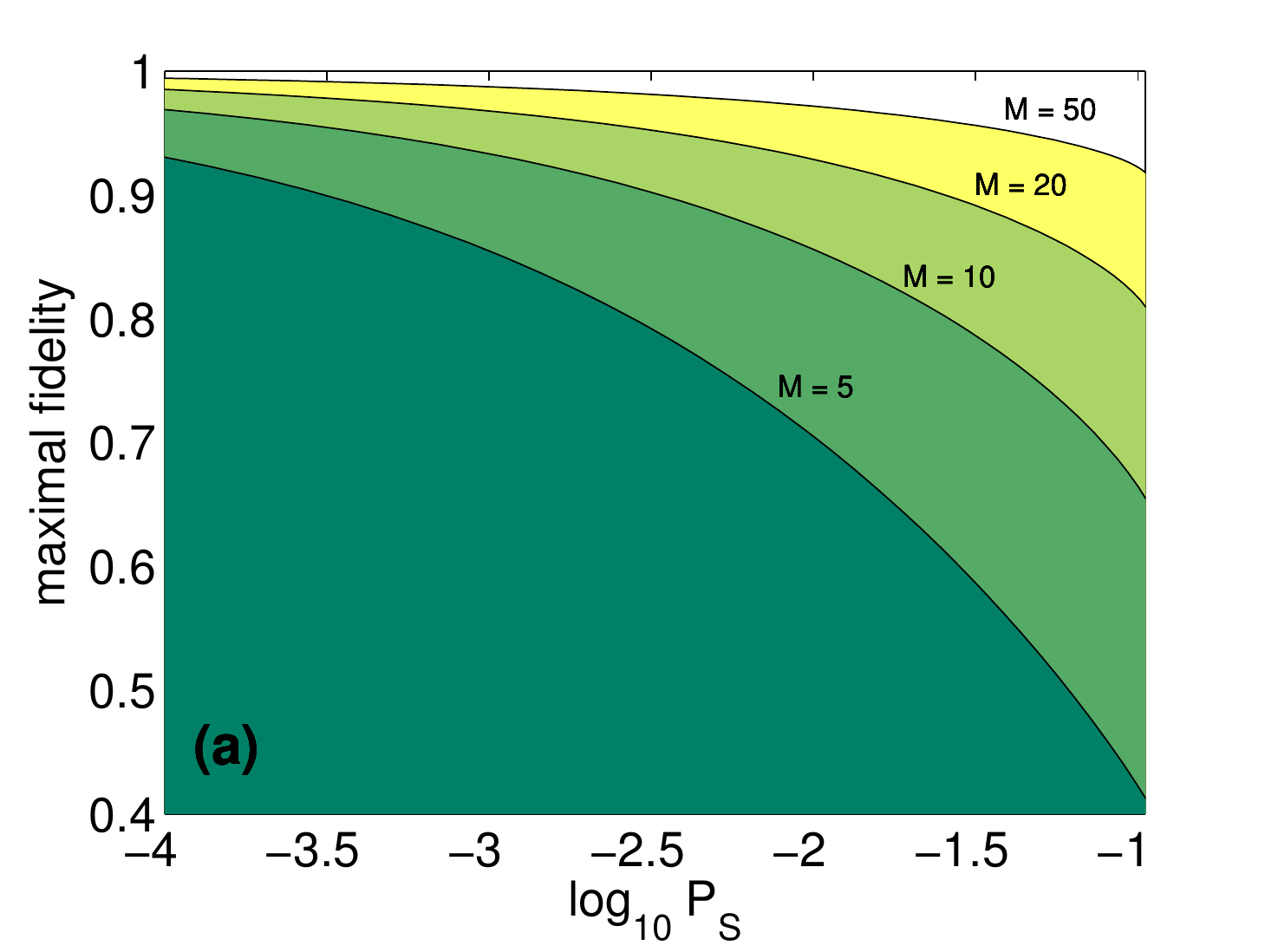}
\includegraphics[width=0.45\textwidth]{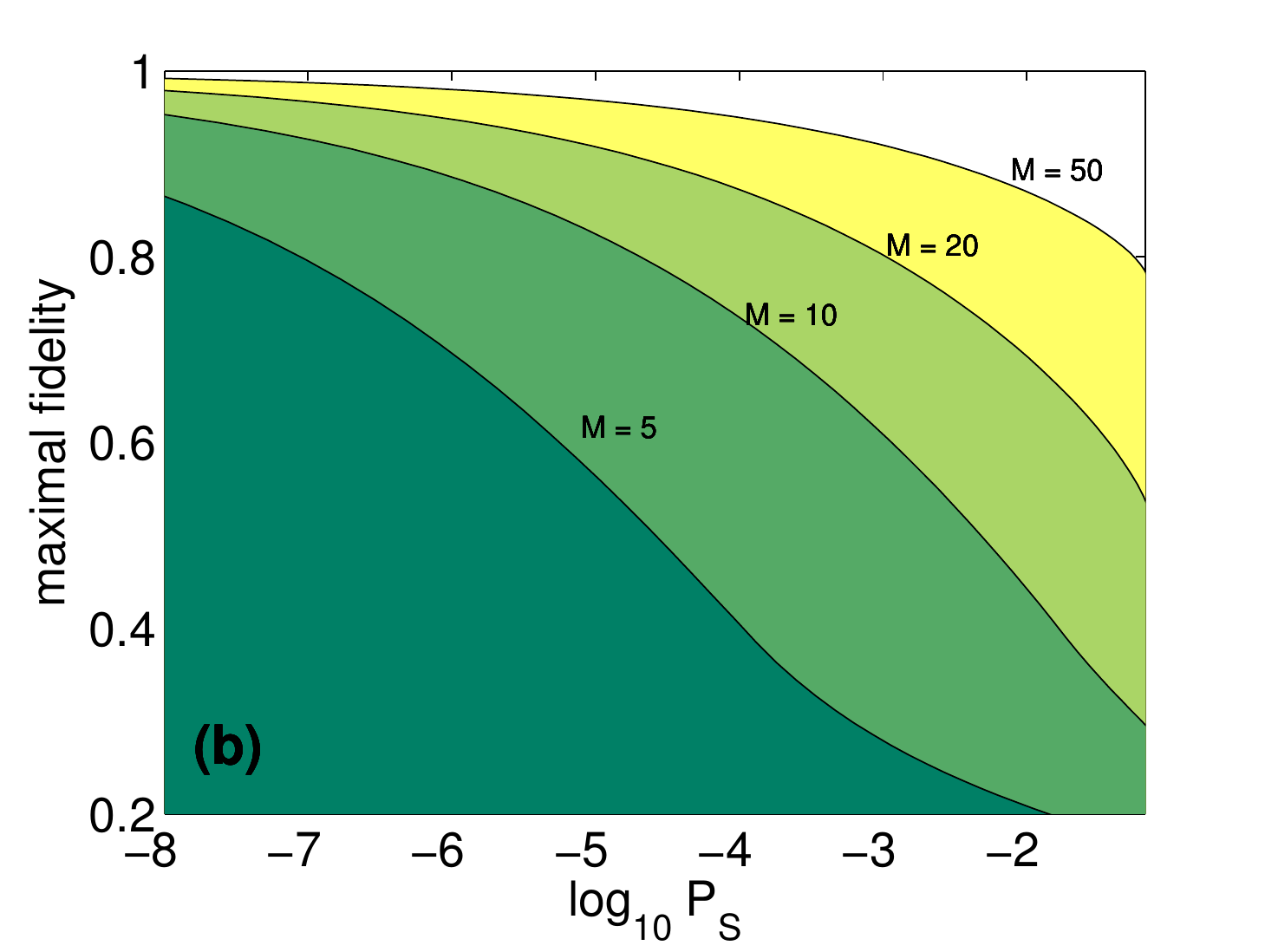}
    \caption{Maximal fidelity $F_{\mathrm{id}}$ of states $|3\rangle$ (a) and $|5\rangle$ (b) prepared by MSPDs with $M = 5, 10, 20, 50$.  }
    \label{fig_benchmarks}
\end{figure}

The biggest drawback of the direct approach is that the bound is optimized over the entanglement of the resource state. Failure to overcome the value for a given MSPD can be therefore caused by lower quality of the detector, but also by inability to experimentally prepare the most suitable two mode squeezed states. This issue can be avoided by refining the benchmarking procedure and compare not only the success probabilities, but also the fidelity of the state, which can be obtained, for example, by quantum state estimation. For this, however, the MSPD bound is represented not by a single value, but by a trade-off between the two quantities. Sample benchmarks for preparation of quantum states $|3\rangle$ and $|5\rangle$ are plotted in Fig.~\ref{fig_benchmarks}. Any experimental result above the line formed by the trade-off then confirms superiority of the tested detector over the respective MSPD.

The benchmarks introduced so far are relevant for PNRDs with high quantum efficiency experiments capable of producing entangled states with high purity. When this is not the case, the benchmarks are still valid, but may be too strict to be surpassed. \red{However, in these situations it is possible to derive new, more forgiving, benchmarks that already take the lowered detector efficiency $\eta_M<1$ in the account, similarly to \cite{PhysRevA.88.043827}. Such benchmarks can be obtained by straightforward numerical evaluation of the fidelity with respect to probability of success in a system with detector efficiency $\eta_M$. However, there is another issue of reduced purity of the shared entangled state. Such states can be equivalently represented by pure entangled states with losses in both modes of the state. The losses of the measured mode can be subsumed into the detector efficiency $\eta_M$ and treated in the same way. The losses in the state preparation mode, represented by reduced efficiency $\eta_S$, can be resolved in a similar manner, by deriving new benchmarks with values of $\eta_S$ and $\eta_M$.}

\red{Another way to approach the issue of reduced efficiency $\eta_S$ is by numerically compensating the losses on the measurement data, which is a tool occasionally employed in analyzing experimental results \cite{PhysRevLett.112.033601, PhysRevLett.115.023602}. In our case, in which data is in the form of photon number distribution, the losses can be compensated by the inverse decoherence map \cite{PhysRevA.77.032302}}
\begin{equation}\label{inverselosses}
     \mathcal{L}_{\eta}(\rho) = \sum_{n=0}^{\infty} (-1)^n \hat{L}_n \rho \hat{L}_n^{\dag},\quad \hat{L}_n = \frac{\eta^{-\hat{a}^\dag\hat{a}/2}}{\sqrt{n!}} \left(\frac{1-\eta}{\eta}\right)^{n/2}\hat{a}^n.
\end{equation}
\red{that subtracts the loss from the state and reconstructs it into the form it would have before the lossy channel with transmission coefficient $\eta$. The map is a direct inversion of decoherence map (\ref{losses}), it is trace preserving, but it is not physical. When it is applied to a state that is not produced by losses it can lead to unphysical density matrix. For example, when applied to pure photon number state that is not the vacuum state, the transformed density matrix will always have some negative diagonal elements. If we know the losses $\eta_S$, we can apply the operation directly, but in that case it is better to directly derive the benchmarks for the specific level of losses. The benefit of using the inverse map is that it can be applied even if the state preparation efficiency $\eta_S$ is unknown. For any quantum state $\rho_S$ we can define the idealized fidelity}
\begin{equation}\label{Fidid}
    F_{\mathrm{id}} = \max_{\eta} \langle n|\mathcal{L}_{\eta}(\hat{\rho}_S)|n\rangle,
\end{equation}
\red{where the maximum is taken over all $\eta$ that produce a physical state, which avoids the explicit dependency on $\eta_S$. Value (\ref{Fidid}) represents the maximal population of the given Fock state that can be, with regards to some level of losses, compatible with state $\rho_S$. This value can then be compared to the respective benchmarks. In order to assure the measured states are not given an unfair advantage, we need to apply the inverse decoherence map also to the photon number distributions obtained during generation of the benchmark values. We have numerically derived the benchmarks for idealized fidelity and compared them to the benchmarks for fidelity. For the cases we considered, preparation of photon number states $|3\rangle$ and $|5\rangle$ with $0.1 \le \eta_M \le 1$, the benchmarks for the idealized fidelity are identical to benchmarks for the fidelity, as long as $F \ge 0.4$. When the fidelity is lower, which happens in the regime with high entanglement of the shared state, low number of $M$ of detectors, and low detector efficiency $\eta_M$, the benchmarks for the idealized fidelity are higher, which makes them more difficult to pass. Fig.~\ref{fig_benchmarks} actually shows these more strict benchmarks. In experiment, this regime can be avoided by limiting the energy of the shared entangled state and focusing the comparison on events with low probability of success.}

\section{Conclusion}
To fully describe a realistic photon number resolving detector we need to estimate its numerous POVM elements. These elements depend both on the detector's quantum efficiency and on its inherent response to different photon number states. The most serious drawback of this approach is that each of these, in principle, infinitely many elements is an operator of infinite rank, which clashes with the practical need for simple characterization.

We have shown that there is a single parameter that can be used as a short descriptor of the overall behavior of an indeterminate photon number resolving detector. It is the number of ideal on-off detectors that need to be used in a chosen practical scenario to obtain the same results as the tested PNRD. We suggest that the most suitable scenario is that of conditional preparation of photon number states, in which we can evaluate not only the quality of the detectors but also their compatibility with the quantum optics toolbox.

The benchmarking itself can be realized in three steps. In the first one we can use the unknown detector as both the conditioning and the verification measurement. We experimentally try to maximize the joint probability of success and compare it to respective benchmarks for different MSPDs. In the second step the unknown detector is used only for conditioning and the prepared state is estimated in order to obtain fidelity. The fidelity and success probability are then compared to the benchmarks. Finally, estimation can be used to obtain the full photon number distribution which can be then employed to compensate for the inherent losses in the system.

\section*{Funding}
We would like to acknowledge grant GA18-21285S of the Czech Science Foundation. J.P and L. L. would also like to acknowledge internal projects of Palack\'y University IGA-PrF-2019-010 and IGA-PrF-2020-009. We also acknowledge national funding from the MEYS and the funding from European Union’s Horizon 2020 (2014-2020) research and innovation framework programme under grant agreement No 731473 (project
8C20002 ShoQC). Project ShoQC has received funding from the QuantERA ERA-NET Cofund in Quantum Technologies implemented within the European Union’s Horizon 2020 Programme.

\section*{Disclosure}
The authors declare no conflicts of interest.

\section*{Appendix A: Proof that MSPD approaches PNRD}
This limit hold when $\lim_{M\rightarrow \infty} p(n|k) = \delta_{n,k}$.
First, let us show that $\lim_{M\rightarrow \infty}  p(n|n) = 1$. Directly evaluating  (\ref{POVM_APDprob}) we get
\begin{align}
    \lim_{M\rightarrow\infty} p(n|n) = \lim_{M\rightarrow\infty} \frac{M!}{M^n n! (M-n)!}
    \sum_{l= 0}^{n-1} \left( \begin{array}{c}
                               n \\
                               l
                             \end{array}\right)
                             (-1)^l (n-l)^n \nonumber \\
    = \lim_{M\rightarrow\infty} \frac{\mathcal{P}_M(M)}{M^n\mathcal{P}_{M-n}(M)}  \sum_{l= 0}^{n-1} \frac{(-1)^l (n-l)^n}{l! (n-l)!}
\end{align}
where $\mathcal{P}_m(x)$ denotes $m$-th order polynomial of variable $x$. The whole limit is then equal to 1, because the fraction of the polynomials scales as $\frac{M^n}{M^n}$ and the sum is equal to 1 for all $n$.

Similarly, for $\lim_{M\rightarrow \infty}  p(n|k) = 0$ when $n\neq k$,
\begin{align}
    \lim_{M\rightarrow\infty} p(n|n+z) = \lim_{M\rightarrow\infty} \frac{M!}{M^{n+z} n! (M-n)!}
    \sum_{l= 0}^{n-1} \left( \begin{array}{c}
                               n \\
                               l
                             \end{array}\right)
                             (-1)^l (n-l)^{n+z} \nonumber \\
    =  \lim_{M\rightarrow\infty} \frac{\mathcal{P}_M(M)}{M^{m+z}\mathcal{P}_{M-n}(M)}  \sum_{l= 0}^{n-1} \frac{(-1)^l (n-l)^{n+z}}{l! (n-l)!},
\end{align}
which is equal to zero because the sum is a finite number while the limit of the fraction of polynomials is vanishing.

%\bibliography{manuscript_ref}

\end{document}